# Supporting Knowledge and Expertise Finding within Australia's Defence Science and Technology Organisation

Paul Prekop
*DSTO Fern Hill, Department of Defence, Canberra ACT 2600*
*paul.prekop@dsto.defence.gov.au*

## Abstract

*This paper reports on work aimed at supporting knowledge and expertise finding within a large Research and Development (R&D) organisation. The paper first discusses the nature of knowledge important to R&D organisations and presents a prototype information system developed to support knowledge and expertise finding. The paper then discusses a trial of the system within an R&D organisation, the implications and limitations of the trial, and discusses future research questions.*

## 1. Introduction

This paper describes work undertaken to support knowledge and expertise finding within Australia's Defence Science and Technology Organisation (DSTO). DSTO is a government funded research and development (R&D) organisation, with a very broad, applied R&D program focused primarily within the defence and national security domains. DSTO employs approximately 1900 engineers and scientists across a wide range of academic disciplines (about 30% of staff hold PhDs), within seven sites throughout Australia.

Like most other large R&D organisations [1, 2] and professional services firms, DSTO is a project-centric organisation; projects are formed to address specific questions or problems, or to develop specific products. The nature of the outcomes of the projects undertaken by DSTO varies considerably, and can range from academic papers and technical reports, through to prototype and working system development, and to professional services and consulting engagements.

The work described in this paper is part of an ongoing knowledge management improvement program aimed at exploring:

- Methods to allow staff to build and maintain wide and detailed awareness of DSTO's past, current and planned projects;
- Methods to enable staff to locate other staff with relevant skills, interests, abilities or experience;
- Low cost (in terms of time and effort) methods to support the development of communities of interest, and less-formal collaboration and sharing within the organisation;
- Organisational cultural and behavioural issues that may act as barriers to effective knowledge and expertise sharing.

A prototype information system, the Automated Research Management System (ARMS), was developed to explore approaches to addressing these issues.

Section 2 discusses the nature of knowledge and knowledge management within the R&D environment, and describes the types of support for knowledge and expertise-finding needed within organisations such as DSTO. Section 3 describes ARMS and how it supports knowledge and expertise finding within DSTO. Section 4 outlines the ARMS trial and trial methodology, and Section 5 discusses the results of two studies undertaken as part of the ARMS trial. Finally, Section 6 discusses the implications and limitations of the work undertaken so far and describes potential areas for future work.

## 2. The Nature of Knowledge and Expertise within R&D Organisations

### 2.1. Theoretical Background

The main theoretical idea underpinning this work is that the knowledge important to an organisation, or that makes it unique or gives it a competitive advantage, is embedded in key elements that make up the organisation [3–5].

According to [3], this knowledge is embedded in three key organisational elements – the members of the organisation, the tools used within the organisation, and the tasks performed by the organisation. For many organisations, organisationally important knowledge is embedded within skills, experiences, expertise and competencies of the individuals that make up the organisation [1, 3, 6]. This is particularly true for R&D organisations [1] and other professional services organisations [5]. As well as people, significant organisational knowledge is embedded within the tools the organisation uses, including specialised physical hardware used as part of a manufacturing process, for example, through to conceptual or intellectual tools such as consulting or analysis frameworks [3, 7]. The third key element identified by [3] is the tasks performed by the organisation. Tasks reflect an organisation's goals, intention and purpose [3, 5, 7].

For R&D, engineering and other professional services organisations, key knowledge is also embedded within the products or other kinds of outcome the organisation produces. The development of products or other kinds of outcome uniquely combines together the organisation's staff, tools and tasks to address a particular question or problem, or to develop some kind of product, and can be seen as uniquely embedding the application of the organisation's collective knowledge, skills, experiences and expertise within a particular domain, to address a particular question or problem or to develop some kind of product [1, 7, 9, 10].

As discussed in [11], knowledge management is centred on two, potentially limiting, philosophical foundations. The first is the idea that tacit and explicit knowledge are two distinctly different forms or types of knowledge, rather than simply being a dimension along which all knowledge exists. The underlying assumption that tacit and explicit knowledge are different leads to the conclusion that a key goal of knowledge management is the codification of tacit knowledge into explicit knowledge [12]. However, as [11] points out, not all knowledge can (or should) be codified, and any knowledge management approaches that rely on the codification of knowledge are likely to fail. The second philosophical foundation that knowledge management rests on is the *data – information – knowledge* continuum: the idea that information is in some way better data and that knowledge is in some way better information. As discussed in [11], this leads to knowledge management approaches that focus only on capturing and managing some form of codified knowledge [13], while ignoring data and information that could provide equal or even greater value to users.

However, the view that knowledge important to an organisation is embedded in the key elements that make up the organisation potentially provides an approach to knowledge management that doesn't rest on these two potentially limiting foundations. The focus of a knowledge management approach that accepts the embeddedness of knowledge as important becomes one of finding and devising methods, systems and approaches that in some way index and expose the core entities within the organisation that hold the embedded knowledge, rather than focusing on codification and managing the codified knowledge. The importance of the knowledge embedded in the different organisational entities will vary with the nature of the organisation and the nature of the work performed by the organisation. The following section discusses the kinds of knowledge important within an industrial R&D organisation, and the kinds of entities the knowledge is likely to be embedded within.

## 2.2. Knowledge and Expertise within R&D Organisations

For industrial R&D organisations (and many other knowledge intensive firms [14]), the key knowledge that makes the organisation unique is embedded in the experience, expertise and competencies of the engineers and scientists that make up the organisation and the organisation's collective project history – the past products the organisation has developed, or the past problems or questions it has addressed [1, 10].

Within most industrial R&D organisations, the products developed or projects undertaken require the application of collective individual knowledge, skills, experience and expertise in unique ways. As a result, products and projects can be seen as an embedding the application of the organisation's collective knowledge, skills, experience and expertise, within a particular domain, to address a particular question or problem [1, 8].

The information and knowledge created as part of past projects can provide important insights into finding solutions to current problems, or gaining an understanding of how similar problems have been solved in the past. Project histories can also support problem reformulation, and can offer some help in validating proposed solutions [15]. Past projects are also important because they can provide a link back to the staff who contributed. This in turn can provide valuable insight into the knowledge, experience and expertise that individual staff may have [1, 8, 15]. Colleagues act not only as important sources of information, and pointers to other sources of information [16], but most importantly they provide an interactive think along function [17, 18].

As discussed previously, the goal of this work was to develop ways of improving knowledge and expertise finding. The approach taken by the work described in the following section was to develop an information system (ARMS) to act as a rich, interactive model of the embedded knowledge within the organisation – in particular, the current and past projects within the organisation, and the expertise, skills and experience of the staff that make up the organisation.

## 3. The ARMS Prototype

The Automated Research Management System (ARMS) is a prototype, web based, information system developed to explore approaches to supporting knowledge and expertise finding within DSTO.

ARMS holds information about the key R&D entities relevant to DSTO – staff and projects, as well as formal and informal project outputs (academic papers, technical reports, design documents, data collections, and so on). Within ARMS, these entities are organised around the organisation's hierarchical structure, and around Themes – collections of taxonomic descriptors used to describe the client and scientific domains that DSTO works within. The key R&D entities and their relationships are shown in Figure 1.

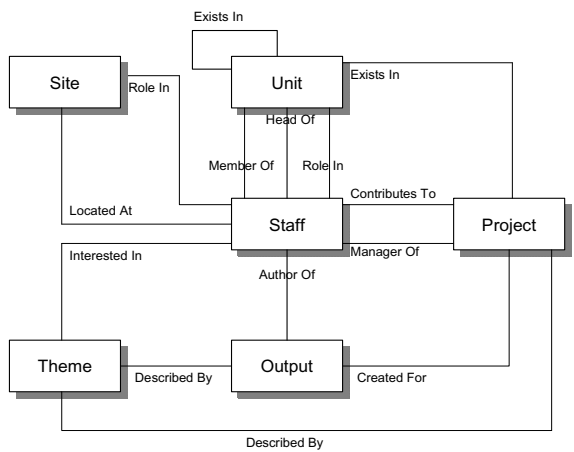

**Figure 1. Conceptual domain model**

### 3.1. The Users' Perspective

As discussed in Section 2, the key knowledge, experience, expertise and competency of an industrial R&D organisation exists in its project history and the collective expertise, skills, experience and knowledge of its scientists and engineers. ARMS supports knowledge and expertise finding by exposing both projects (and the outputs associated with projects) and staff as richly interlinked first class objects (unlike many similar systems [19], where staff are either not included, or simply included as author labels associated with the entities held by the system).

From a user's perspective, ARMS exists as a collection of dynamically generated web pages, with each R&D entity having its own web page that pulls together all the information related to that entity.

Staff pages contain basic staff information, including: contact information, site location and organisational affiliation, and descriptions and links to the projects and project outputs the staff member has contributed to. This information is drawn from existing organisational information systems. Staff pages can also contain optional information, including staff descriptions of current and past project work, staff photo and basic biographical information, and current interests and work.

Project pages contain project descriptions (abstract, overview, background, themes, etc), project milestones, planned deliverables, information about the project's relationship to other past current and planned projects, and the project's status. Project pages also contain information and links to the staff that have contributed to the project, as well as information about and links to the project's outputs. All this information is obtained from existing information systems. End users are also able to add richer project description information, as well as any kinds of additional outputs to the project's home page.

Output pages contain basic metadata describing the output (title, abstract, publication details, document type and so on) and the documents that make up the output (for example MS-Word files, data files, image files, etc). Output pages contain information and links to the staff who contributed to the output, as well as information and links to the project the output was developed for. Output information contained within ARMS is extracted from an existing publication management system. End users are also able to add additional project outputs to any project page.

ARMS provides multiple entry points into the data. Users can access staff, project and output entities directly via the staff, project and output browsing pages. Each of these pages provides filterable lists of the staff, project and output entities contained within ARMS.

In addition to the browsing pages, users can also enter ARMS via a representation of the organisation's structure (the Unit entities shown in Figure 1), or via descriptions of the organisation's R&D program (the Themes entities shown in Figure 1).

Each DSTO unit has a corresponding unit page that holds basic contact information for the unit, the unit's head, administrative contacts for the unit, and the unit's structure (for example the groups within a branch), the staff who are members of that unit, and the tasks associated with that unit. This information is extracted from several different existing information systems. In addition, end users can optionally insert additional information describing units. Units can be directly entered via the unit browse page, an interactive form of the organisation's hierarchical structure.

DSTO's R&D program is represented as a collection of Themes, a combination of taxonomy descriptors (based around DEFTEST [20]) that cover the core science and technology (S&T) areas of DSTO, and taxonomy descriptors that cover the key client areas DSTO works within. S&T and client Themes are used to describe the staff, project and output information held within ARMS. Each Theme has an automatically generated home page that aggregates all the projects, staff and outputs described by the Theme. In many ways the Theme home pages can be seen as aggregating everything the organisation 'knows' about a particular Theme area – that is all the staff, projects and outputs related to that Theme. Themes can be directly entered via the Theme browse page, an interactive and structured collection of the Themes held by ARMS.

An important part of ARMS is rich hyperlinking between the various entities that contextualises the information held by ARMS. Staff, for example, are contextualised by the projects they contribute (or have contributed) to and the outputs they have contributed to. Projects are contextualised by the staff that contribute to them, and the part of the organisation they were performed by. The intent of the contextualisation is to allow users to infer richer meanings based on explicit relationships present in the data. The utility of this approach, especially in terms of expertise finding, is discussed in Section 5.2.

In addition to the browsing functions, ARMS also includes a search function that support free text searching over all the information held by ARMS (fields associated with each entity as well as full document text), field searching (for example, searching explicitly by unit name, or project number), as well as Theme searching. The different search types can be combined with Boolean operators (AND and OR) to form complex queries.

### 3.2. Technical Perspective

From a technical perspective, ARMS consists of a centralised repository that holds information extracted from existing corporate information systems, as well as a small amount of information specifically created to support ARMS. Access to the information and functions provided by ARMS is via a web based interface that supports the functions described in Section 3.1, as well as a Web Services[1] interface that provides dynamic programmatic access to the core functions and information (see Figure 2). The Web Services interface was provided to support dynamic access to by ARMS by specialised applications, such as collaborative Microsoft SharePoint portals [22] and other specialised web sites.

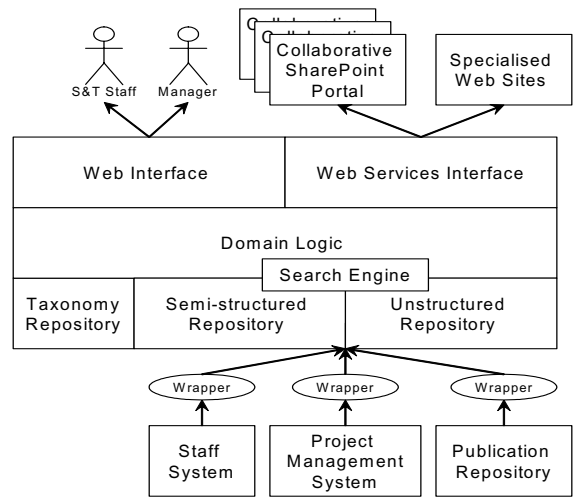

**Figure 2. ARMS logical architecture**

Almost all of the information used by ARMS was drawn from existing corporate information systems as well as less-formal corporate information collections such as spreadsheets and intranet sites. The re-used corporate information generally needed to be cleaned and reformatted before it could be used. Most of the data cleaning issues encountered were common to other data warehousing projects [23], and included a lack of common record identifiers across the different systems, conflicting data values across the different systems, missing and erroneous data, and name and type conflicts. The data cleaning and re-formatting functions were encapsulated in series of customer wrappers that were developed for each of the core corporate applications and other corporate information sources.

As discussed in Section 3.1, ARMS provided a search function over all the information held. The search engine combined free text searching over the unstructured information held by ARMS (generally reports, papers, presentation and other project outcomes) with searching over the structured and semi-structured information held. Free text search was provided by a commercially available text search engine; searching over structured and semi-structured information was provided by a custom developed search engine.

---

[1] An overview of Web Services can be found in [21].

## 4. The ARMS Trial

To validate the concepts underpinning ARMS and its utility in supporting expertise and knowledge finding within DSTO, a prototype version of the system was evaluated within two DSTO divisions from June until November 2004.

The ARMS prototype (as described in [24]) was fully developed and fully populated with relevant staff, project and output information. The data held by ARMS was kept up-to-date throughout the trial period. During the trial period ARMS was made available to staff within the two divisions selected (Division A and Division B). These divisions were selected because together they reflect a good cross section of the staff, organisational structures, and research programs within DSTO.

Division A was split across five sites, and generally had a multi-disciplinary, professional services focus. At the time of the trial it contained approximately 80 staff.

Division B was split across three sites, and generally had a computer science/software engineering base, with a strong R&D focus. At the time of the trial, it contained approximately 110 staff.

**Table 1. ARMS Trial Studies**

| Study | Date | Collection Methods |
|---|---|---|
| Study One: Organisational Wide Focus Groups | May | Focus Groups |
| Study Two: Stake-holder Interviews | June/July | Unstructured and semi-structured interviews |
| Study Three: Concept and Implementation Survey | July | Survey questions |
| Study Four: Project Seeking Experiment | September | Data seeking experiment |
| Study Five: Utility of Usage Study | November | Semi-structured interviews and survey questions |

Over the trial period, five different studies were undertaken (see Table 1). Each of the studies aimed to explore the utility of ARMS from various perspectives. This paper discusses two of the key studies, Study Three and Study Five, both of which measured the utility of ARMS from the perspective of R&D staff.

The Concept and Implementation Survey (Study Three) was undertaken after ARMS had been available in the two trial divisions for approximately 1½ months. The survey was sent to all staff in the two trial divisions who had used ARMS at least once during the trial period. Of the 75 divisional staff who had used ARMS at least once, 23 staff responded.

Respondents were spread across three sites; 40% from Site A, 52% from Site B, and 8% from Site C. The respondents represented a good cross section of the organisation's management structure, with 17% of respondents holding organisational unit management positions (group or branch), and 40% of respondents having project management responsibility. Overall the sample reflected a good cross-section of the organisation, with slight over-sampling of respondents from Site B. The results of this study are discussed in Section 5.

The Utility of Usage Study (Study Five) was undertaken towards the end of the ARMS trial. The most frequent users of ARMS were identified, and invited to participate in the study. Of the users selected, 10 agreed to participate. Study respondents were spread across three sites; 70% from Site A, 20% from Site B and 10% from Site C. Around half of the respondents held project management responsibility. Overall, the sample in this study reflected a good cross section of the organisation, except for the lack of respondents holding organisational unit management positions, and an over-sampling of respondents from Site A.

Study Five was run as a set of survey questions and semi-structured interviews. Participants were asked to list the functions they used ARMS to perform over the previous month. The interviews discussed how participants used ARMS, its utility to them, and how ARMS compared to alternative approaches they may have tried. The interviews were recorded, and the recordings transcribed. The interviews ran for an average of 50 minutes each.

## 5. Results and Discussion

### 5.1. ARMS Usage

Studies Three and Five recorded how ARMS was used by participants over the previous month. Table 2, below, lists the functions ARMS was used to perform and the percentage of users who used ARMS to perform the listed function.

One of the most striking features of Table 2 is the dramatic changes in some of the ways ARMS was used over the trial period. (Study Three was undertaken after ARMS had been available for 1½ months; Study Five was undertaken toward the end of the 6 month trial).

**Table 2. Overall ARMS Usage**

| Use of ARMS | Study Three | Study Five |
|---|---|---|
| 1. Exploring the system; for example seeing what information it holds and exploring how I could use it. | 96% | 20% |
| 2. Building an awareness of DSTO's projects and staff; for example seeing what work a group is doing, or what work is going on within a Work Area, or seeing what a particular person has been working on recently. | 61% | 10% |
| 3. Finding out about a particular project; for example who is working on it, who the project manager is, or finding the various papers, reports and other outputs associated with the project. | 56% | 40% |
| 4. Finding out about a person; for example their contact information, what they are working on, what papers and reports they have written or what outputs they have been involved in, or finding out about their interests or experience. | 52% | 50% |
| 5. Finding out about a particular research area; for example who is interested in the area, which tasks contribute to the research area, or what work has been produced in the research area. | 35% | 20% |
| 6. Finding out about a particular organisational unit; for example who is in the unit, what tasks the unit is responsible for, or finding contact information for the unit. | 30% | 0% |
| 7. Finding out about a particular Client Area; for example who is working in, which tasks contribute to the Client Area, or what work has been produced for the Client Area. | 22% | 0% |
| 8. Finding a particular formal or informal output. | 13% | 30% |
| 9. Searching for a person with particular skills, interests, experience or abilities. | 13% | 10% |
| 10. Other | 4% | 0% |

The largest change was the drop in exploratory use of ARMS between the two studies. This change is likely to be a result of users building an understanding of the functions and features of ARMS as the study progressed. Once an understanding of ARMS (based on exploring the system) was developed they either moved to non-use, if they felt ARMS provided no value, or they moved to using specific features and functions of ARMS.

Using ARMS to build and maintain an awareness of the work being performed within DSTO (Question 2), and using ARMS to find out about particular organisational units (Question 6) and client areas (Question 7) also dropped over the study period. The drop in using ARMS to perform these functions is likely to be due to two key factors. The semi-structured interviews undertaken as part of Study Five revealed that awareness (Question 2) is something that is built for a specific purpose, for example moving to a new organisational unit or moving into a new research area, and then is maintained by actively being involved in the organisational unit or research area. Once an overall awareness has been built, it is maintained by more direct information seeking activities, for example finding out about a staff member, or a project, or hunting up specific project outputs.

The changes in the results for Questions 6 and 7 are due to similar reasons, with participants initially using ARMS to build a general awareness of organisational units and client areas, and then maintaining that awareness by more direct information seeking methods.

The second factor likely to affect the use of ARMS to build and maintain awareness is a fundamental limit inherent within the ARMS trial. As discussed previously, ARMS was fully populated and maintained with data drawn from only two trial divisions, not the whole organisation. As a result the value of ARMS in providing a rich awareness to participants was limited to only the two trial divisions. This limitation and its likely impact on the overall trial result is discussed in more detail in Section 6.1.

### 5.2. Finding People, Projects and Outputs

One set of functions that ARMS was regularly used to perform over the study period was finding people, projects and outputs (Questions 3, 4 and 8 respectively).

As shown by Table 2, finding people was constantly the most frequently used function of ARMS. The semi-structured interviews undertaken as part of Study Five revealed that participants who used ARMS to find people were generally looking for colleagues who have similar research interests or worked in similar areas. The semi-structured interviews revealed that users who employed ARMS in this way, were, in general, seeking out connections with other colleagues in order to share ideas, discuss problems/issues, have work reviewed, gain insights and Proceedings of the 40th Hawaii International Conference on System Sciences - 2007different perspectives on a common/shared problems, and perhaps even re-use and build on the work already completed by colleagues.

These findings confirm much of the previous research (described in Section 2) describing the information seeking behaviours and needs of scientists and engineers.

Overall, participants reported that the information available via ARMS in most cases allowed them to identify colleagues they felt would be worth contacting. However, there were some limitations in the data ARMS provided. One key limitation was the scope of the information available via ARMS. As discussed previously, ARMS only contained information for the two trial divisions. This limited the number of colleagues users could learn about. The second key limitation with the staff data available via ARMS was that individual staff expertise, experience and research interests, in general, could only be derived by understanding the context surrounding the staff member (this is discussed in Section 5.3). While in most cases the surrounding context provided users with sufficient information to infer a staff member's expertise, experience and research interests, it did mean that users were generally unable to directly search for staff with particular expertise, experience and research interests but instead had to search for staff via outputs or projects, and infer staff relevance based on the relationships between staff and the projects and outputs they contributed to.

An interesting question explored as part of the semi-structured interviews was the relationship between the kinds of information available via ARMS, and the kinds of information available via the user's social networks. Most respondents acknowledged the importance their social network plays in being able to find colleagues with specific expertise, experience and research interests. However, they generally found that there were limits in the coverage of their social network; in particular many respondents felt their social network often didn't extend into other organisational sites, or into other organisational units. In comparison, they felt that the information held by ARMS was more complete and offered greater coverage of all parts of the organisation, and they viewed ARMS as a way of filling gaps in their social network.

The second key group of functions regularly used throughout the trial was finding projects and outputs. The semi-structured interviews revealed that when searching for projects and outputs, participants were generally looking either to find relevant and useful colleagues, or were explicitly looking for information related to past projects. As discussed in Section 2, for engineers in particular, descriptions of past projects and the formal and informal products of projects (reports, papers, designs, data sets, meeting minutes, and so on) are important because they offer insights and approaches to solving past problems that may be useful for solving current problems. The semi-structured interviews revealed that participants were using ARMS in this way.

As well as ARMS, participants could obtain project and formal publication information from two existing organisational information systems, a Project Management System (PMS) and a Publications Repository (PR). The basic project and publication information held by ARMS was obtained from these two systems, and in most cases ARMS held little additional information. The key value added by ARMS was the rich contextualisation of the information, together with existing staff information to provide multiple entry point into the information sought, and to allow rich relationships and deeper meaning of the information to be inferred. This is discussed in more detail in the following section.

### 5.3. Contextualisation and Navigation

As shown by Figure 1, the information contained within ARMS is richly interlinked. The rich interlinking helps to contextualise this information, making it easier for ARMS users to develop a deeper understanding of the information held within ARMS, as well as providing users with multiple entry points into the information, and a navigation model drawn from the users' domains.

The semi-structured interviews undertaken as part of Study 5 showed that by knowing, for example, the background of key staff involved in a project, users are able to infer more about the likely directions, perspectives or methodologies a project may use. Or by knowing about the project that created a particular output, users are able to build a richer understanding of the output's meaning.

The rich interlinking within ARMS also provided users with multiple entry points into the information held by ARMS, and provided them with natural navigation paths drawn from their domain. The semi-structured interviews revealed that almost all users reported that the richly contextualised nature of ARMS helped them navigate ARMS, allowing them to enter ARMS from many different directions, often using only incomplete information as a starting point.

For example, several users reported browsing ARMS by starting with a vague awareness that a staff member may be doing work that might be interesting or relevant to them. By using ARMS, they were able to move from the individual's staff page to their project pages – to find out about the work being performed. From project pages, they would move to output pages, to gain a deeper insight of the work being performed, or they would move to Theme pages, to find related work, staff or outputs. Other users reported a similar approach using projects as a starting point to find people or to find other related projects.

This form of navigation also allows for far richer forms of accidental information discovery [25], and provides users with the ability to develop rich mental models of the organisation's past, planned and current work program.

### 5.4. Cultural Impact

As discussed in Section 1, one of the goals of the work described in this paper was to address potential barriers to information sharing within DSTO, including behavioural and cultural issues, as well as a lack of low cost methods, systems or processes to support information sharing within DSTO. Both Study Three and Study Five attempted to measure the likely influence ARMS would have on these barriers. The results are shown in Table 3; questions were measured on a 7 point, end anchored scale, with 1 meaning *strongly disagree*, and 7 meaning *strongly agree*.

**Table 3. Cultural Impact**

| Question | Study Three | Study Five |
|---|---|---|
| If ARMS held information about everyone in the DSTO and all DSTO's past, current and planned work: | | |
| 1. It would be easy to find out what is going on in DSTO? | $\bar{x}$ = 5.55<br>$s$ = .80 | $\bar{x}$ = 6.11<br>$s$ = .78 |
| 2. It would encourage people to work together rather than compete with one another? | $\bar{x}$ = 4.61<br>$s$ = 1.16 | $\bar{x}$ = 4.58<br>$s$ = .88 |
| 3. It would encourage people to share the information they have, or their knowledge and expertise? | $\bar{x}$ = 4.83<br>$s$ = 1.07 | $\bar{x}$ = 4.67<br>$s$ = .87 |

The most interesting insight from Table 3 is the different perceptions respondents had of the role ARMS could play in lowering the cost of sharing within the organisation (Question 1), versus the ability of ARMS (or any information system) to influence the behavioural and cultural issues that affect information and knowledge sharing within the organisation (Questions 2 and 3).

As discussed previously, almost all of the information included within ARMS was drawn from existing organisational information systems. As a result, users were able to build a basic understanding of the past, current and planned work within the organisation, and the skills, interests and experience of staff without requiring the staff described by ARMS to actively make the information available. The strong results for Question 1 across both studies, and the positive insights gathered by the semi-structured interviews, show that, within the trial organisation, the approach of populating ARMS with existing corporate information did lower one of the barriers to sharing by providing a low cost method for 'finding out what is going on in the organisation'.

However, the semi-structured interviews undertaken as part of Study Five revealed that, while being able to find out what was going on in the organisation is a necessary first step towards supporting sharing and collaboration, by itself, it isn't sufficient [14] to significantly impact existing behavioural and cultural issues affecting sharing within DSTO. Many factors affect sharing, including: organisational incentives (for example, rewards, time, encouragement) and disincentives; individual behaviours; the need to derive value from sharing and collaborating; organisational structures; and administrative barriers.

### 6. Conclusions

This paper has described ARMS, an information system aimed at supporting knowledge and expertise finding within DSTO. ARMS was trialled within two divisions of DSTO over a six month period. During this time, five different studies were undertaken; this paper has reported on two of the key studies aimed at measuring the utility of ARMS from the perspective of R&D staff.

The work described in this paper has three key implications. The first is that, within the trial organisation, ARMS provided considerable value in supporting the information and knowledge management needs of R&D staff. In particular, ARMS provided a mechanism by which R&D staff could find colleagues that had similar research interests, or had expertise or insights that could help with a particular problem. ARMS also provided a rich corporate memory function, allowing R&D staff to browse the organisation's previous work. As discussed in the literature (see Section 2), finding staff, and finding previous work are two key information and knowledge needs for staff within industrial R&D organisations.

The rich interlinking of the information held by ARMS helped to improve the usability of the information held by providing multiple entry points into the information, and intuitive navigations paths that matched the user's model of the domain. The rich interlinking also allowed users to see the information held by ARMS in a wider context, and this often enabled them to build a much richer understanding of the information.

By reusing existing corporate information, and not requiring R&D staff to add information to ARMS to expose their skills, expertise or experience, or to expose the past and current work within the organisation, ARMS helped overcome one of the key cultural and behavioural barriers toward sharing within DSTO. While ARMS provided the necessary first step towards improving such sharing, ARMS (or any IT system) is still likely to have a limited impact on entrenched organisational cultural and behavioural barriers [26].

### 6.1. Limitations

The work described in this paper has been applied only within one R&D organisation. While much of the data collected suggests that the information and knowledge seeking needs of DSTO's scientists and engineers is similar to other R&D organisations reported in the literature, the information and knowledge management problems faced by DSTO may be relatively unique, and as a result, the value of an information system like ARMS to other R&D organisations may be different.

A second key limitation of this work was the scope of the information made available within ARMS over the trial period. As discussed previously, only the information related to the two trial divisions was made available within ARMS. The data collected as part of the studies showed that, in many ways, this limited the overall utility of ARMS, especially in supporting information seeking outside of the two divisions.

### 6.2. Future Work

A significant feature the current implementation of ARMS lacks is automatically generated, easy to navigate descriptions that describe the skills, expertise and experience of individual staff. While ARMS users were generally able to infer and derive the likely skills, expertise and experience of individual staff (as described Section 5.3), the lack of automatically derived descriptions of skills, expertise and experience of individual staff limited the ability of users to directly search and browse for staff by skills, expertise and experience. Previous work [27, 28] has shown positive results in deriving descriptions of individual's expertise from descriptions of the work they perform, or the roles they hold. Given that ARMS already strongly relates well described projects and outputs to individual staff, it would be possible to automatically derive or infer descriptions of individual staff expertise from the information already contained within ARMS.

By drawing together staff, projects and outputs, Theme pages act as a potential community of practice hub because they provide wide awareness of an individual's particular expertise or interests, and provide resources relevant to that particular area (projects and outputs and their descriptions). Potentially, the Theme home pages could be expanded to include additional functions to encourage the creation of community – for example, via richer resources, and richer interaction methods (cf. [14]). However, due to time constraints imposed on the development of the ARMS prototype, this approach has not yet been explored.

### 7. Acknowledgements

The author acknowledges the valuable assistance of Dr Mark Burnett, Mr Chris Chapman, Ms Phuong La, and Ms Jemma Nguon, in developing the ARMS prototype, and further acknowledges the assistance of Mr Justin Fidock with some of the evaluation studies.